\documentclass[twocolumn,letterpaper,aps,prc,longbibliography,superscriptaddress,nofootinbib,floatfix]{revtex4-2}

\usepackage[utf8]{inputenc}
\usepackage{xcolor}
\usepackage{indentfirst}
\usepackage{amsmath}
\usepackage[symbol,splitrule]{footmisc}
\usepackage{babel,csquotes,xpatch}
\usepackage{multirow}
\usepackage{dcolumn}
\usepackage{notoccite}
\usepackage{graphicx}
\usepackage{subfigure}
\usepackage{subfiles}

\graphicspath{{./}}

\newcommand{\nuc}[2]{$^{#1}$#2}

\begin{document}

\title{Excitation of the isoscalar giant monopole resonance 
using \nuc{6}{Li} inelastic scattering}

\author{J. Arroyo}\email{jarroyo1@nd.edu}
\affiliation{Department of Physics and Astronomy, University of Notre Dame, Notre Dame, IN 46556, USA}

\author{U. Garg}\email{garg@nd.edu}
\affiliation{Department of Physics and Astronomy, University of Notre Dame, Notre Dame, IN 46556, USA}

\author{H. Akimune}
\affiliation{Department of Physics, Konan University, Kobe 568-8501, Japan}

\author{G. P. A. Berg}
\affiliation{Department of Physics and Astronomy, University of Notre Dame, Notre Dame, IN 46556, USA}

\author{D. C. Cuong}
\affiliation{Institute for Nuclear Science and Technology, VINATOM, 179 Hoang Quoc Viet, Hanoi 122772, Vietnam}

\author{M. Fujiwara}
\affiliation{Research Center for Nuclear Physics, Osaka  University, Osaka 567-0047, Japan}

\author{M. N. Harakeh}
\affiliation{ESRIG, University of Groningen, 9747 AA Groningen, The Netherlands}

\author{M. Itoh}
\affiliation{Cyclotron and Radioisotope Center, Tohoku University, Sendai, 980-8578, Japan}

\author{T. Kawabata}\thanks{Present address: Department of Physics, Osaka University, Toyonaka, Osaka 560-0043, Japan}
\affiliation{Center for Nuclear Study, University of Tokyo, Tokyo 113-0033, Japan}

\author{K. Kawase}\thanks{Present address: Kansai Institute for Photon Science, National Institutes for Quantum Science and Technology, Kyoto 619-0215, Japan}
\affiliation{Research Center for Nuclear Physics, Osaka  University, Osaka 567-0047, Japan}

\author{J.T. Matta}\thanks{Present address: Physics Division, Oak Ridge National Laboratory, Oak Ridge, TN 37830}
\affiliation{Department of Physics and Astronomy, University of Notre Dame, Notre Dame, IN 46556, USA}

\author{D. Patel}
\affiliation{Department of Physics and Astronomy, University of Notre Dame, Notre Dame, IN 46556, USA}

\author{M. Uchida}
\affiliation{Department of Physics, Tokyo Institute of Technology, Tokyo 152-8850, Japan}

\author{M. Yosoi}
\affiliation{Research Center for Nuclear Physics, Osaka  University, Osaka 567-0047, Japan} 

\date{\today}

\begin{abstract}
    The incompressibility of infinite nuclear matter ($K_\infty$) is a parameter in the description of the nuclear equation of state that governs the energy cost associated with density oscillations near the saturation density. The most direct experimental method for studying this property of infinite nuclear matter is to probe the isoscalar giant monopole resonance (ISGMR) in finite nuclei. This work explores the use of \nuc{6}{Li} as a probe to study the ISGMR in several stable nuclei: ${}^{58}$Ni, ${}^{90}$Zr, ${}^{116}$Sn, and ${}^{208}$Pb, as complementary to using inelastic scattering of $\alpha$-particles, which has been used to great effect over the last several decades. Elastic and inelastic scattering data for these targets were collected with 343-\,MeV \nuc{6}{Li} beams. In all nuclei studied in this work, the ISGMR strength distributions extracted from multipole decomposition analyses of the inelastic scattering spectra agree very well with the previously measured ISGMR responses from $\alpha$-particle scattering, establishing the feasibility of employing \nuc{6}{Li} inelastic scattering in investigations of the ISGMR.
    
\end{abstract}


\maketitle

\section*{I. Introduction}

Probing the nuclear equation of state (EoS) provides insight into the bulk properties of nuclear matter. Nuclear incompressibility, $K_{\infty}$, is a parameter that encapsulates the curvature of the EoS for symmetric nuclear matter near the saturation density. This is vital for theories governing the properties of large scale bulk nuclear matter, such as core-collapse supernovae \cite{BETHE1979487,Oertel2017}, and small scale nuclear dynamics, such as high-energy nucleus-nucleus collisions \cite{Bertsch1988}.

The excitation energies of the isoscalar giant monopole resonance (ISGMR) are strongly correlated with the incompressibility of infinite nuclear matter \cite{UGarg}. Measurements of the ISGMR in nuclei far from the valley of stability enable the probing of the incompressibility for neutron matter at sub-saturation density \cite{UGarg}. Exploring isotopic chains far from symmetric nuclear matter would help to constrain the EoS for highly isospin-asymmetric systems, such as neutron stars \cite{1992ApJ...398..569L}. 

The most common method for studying $K_{\infty}$ has been $\alpha$-particle inelastic scattering in finite nuclei, and a robust framework for ISGMR measurements with inelastic $\alpha$-particle scattering has been developed in recent decades whereby the experimental setup, optical model approach, and analysis process have been refined and well replicated. However, studying exotic, shortly-lived nuclei requires switching to inverse kinematics with rare isotope beams. The use of \nuc{4}{He} gas as an active target (AT) is, then, a highly desirable approach to these measurements; the distinct advantages offered by \nuc{4}{He} gas are discussed in some detail by Vandebrouck et al. \cite{Marine1}. Unfortunately, pure \nuc{4}{He} gas sparks in AT chambers at high voltages, leading to the use of deuterium gas in the initial measurements \cite{Monrozeau}. In order to avoid sparking in pure \nuc{4}{He} gas, measurements have also been made with the MAYA AT, with 5\% CF$_4$ gas added as a quencher \cite{Marine1, Marine2, Bagchi}. However, such quenching raises other complications such as contributions to final spectra from reactions with the quenching gas. 

It is, therefore, desirable to explore possible alternative probes for the inverse kinematics measurements. ${}^6$Li might be a viable candidate as a solid target in such experiments, avoiding the aforementioned complexities associated with ATs. Additionally, \nuc{6}{Li} has a low particle emission threshold and breaks down primarily through the $d + \alpha$ channel. This leads to fewer false events following spectrographic filtering. Not only does this aid detector systems by reducing dead times, it also leads to a much lower background contribution when analyzing the inelastic scattering spectra \cite{Zamora}. The improved peak to continuum ratio is certainly most helpful in ISGMR studies. One should bear in mind, however, that the use of solid ${}^6$Li targets would present its own challenges which would need to be met in these measurements.

There has been a handful of notable studies utilizing \nuc{6}{Li} as a probe in forward kinematic experiments at bombarding energies of 26-40\,MeV/u \cite{Eyrich12C, Chen24Mg28Si, Chen2008, Dennert1995} and more recent efforts with 100\,MeV/u \cite{Zamora}. 
In this paper, we report results on the ISGMR strengths for ${}^{58}$Ni, ${}^{90}$Zr, ${}^{116}$Sn, and ${}^{208}$Pb, obtained using \nuc{6}{Li} inelastic scattering at medium energies ($\sim$ 60\, MeV/u). 
This work includes nuclei over a broader range of than previous efforts with \nuc{6}{Li} scattering \cite{Zamora,Chen2008}. Also, the measurements reported here spanned a large enough angular range to carry out a full multipole decomposition analysis (MDA). We find that the extracted ISGMR strength distributions in the nuclei investigated in this work are generally consistent with past results utilizing different probes.


\section*{II. Experimental Details}

 The experimental procedure was, essentially, the same as that employed in the $\alpha$-scattering measurements, the details of which have been provided previously \cite{TaoLi_116Sn,PATEL2013178,Yogesh90Zr}; therefore, only the salient features of the current measurements are provided here. The facilities available at the Research Center for Nuclear Physics (RCNP) at Osaka University, Japan, were utilized for measuring elastic and inelastic scattering spectra for \nuc{58}{Ni}, \nuc{90}{Zr}, \nuc{116}{Sn}, \nuc{208}{Pb} a \nuc{6}{Li} beam of  343\,-MeV energy. The areal densities of the enriched targets employed in the measurements are listed in Table~\ref{tab:table1}.
 
 Inelastically scattered \nuc{6}{Li} were kinematically dispersed according to the ion optics of the Grand Raiden spectrometer such that the \nuc{6}{Li} were incident across a pair of horizontal and vertical position-sensitive multiwire drift chambers with momentum dispersion along the horizontal axis of the detection system. This allows for precise momentum measurements of ejectiles, in addition to simultaneous particle identification from energy deposition in two scintillators following the multiwire drift chambers.

The ion optics of the Grand Raiden operated in the vertical focusing mode focused the \nuc{6}{Li} originating from the target chamber into a coherent peak along the vertical axis of the focal plane. As a result, multi-step processes, such as double-scattering events, the kinematics of which mimic those of true inelastic scattering events and comprise the instrumental background, are not focused but uniformly distributed within the focal plane spectrum, allowing for a clean separation of ``instrumental background'' from the final spectra. A more in depth discussion of Grand Raiden's capabilities can be found in Refs. \cite{GRaiden,Fujiwara2023}.

Inelastic scattering spectra were measured in a forward angular range of 0.5$^{\circ}$ $<\theta_{lab}<$ 9.2$^{\circ}$, with extra time given to extremely forward angles due to the very distinct character of the ISGMR angular distribution in this region.  The inelastic scattering runs spanned an excitation energy range of 5-30\,MeV. Examples of forward-angle spectra are presented in Fig. \ref{fig:fwd_spectra}.

\setlength{\tabcolsep}{14pt} 
\renewcommand{\arraystretch}{1.2} 
\begin{table}[t]
\caption{\label{tab:table1}%
Areal densities for all enriched targets.
}
\begin{ruledtabular}
\begin{tabular}{lcdr}
\textrm{Target}&
\textrm{Areal Density }\\
\colrule
\nuc{58}{Ni} & 1.50\,mg/cm$^2$ \\
\nuc{90}{Zr} & 1.95\,mg/cm$^2$\\
\nuc{116}{Sn}& 4.18\,mg/cm$^2$\\
\nuc{208}{Pb}& 10.0\,mg/cm$^2$\\
\end{tabular}
\end{ruledtabular}
\end{table}

As well, elastic scattering data were obtained over an angular range of 2.5$^{\circ}$ $<\theta_{lab}<$ 30$^{\circ}$. This larger angular range was chosen to ensure extraction of optical model parameters (OMP) to be employed in further analysis. Magnetic field strengths were set such that the horizontal focal plane spanned an excitation energy range of 0-15\,MeV, which was sufficient for obtaining elastic scattering cross sections as well as low-lying excited states cross sections.

\section*{III. Analysis}

The offline data reduction was performed in the ROOT data analysis framework \cite{root}. The data reduction process is identical to that previously described in Ref. \cite{Howard2019}, and will be described here briefly. Particle identification gating was implemented in the scintillator spectra. To contend with the instrumental background, it is worth reiterating that ejectiles scattered before or after the target are defocused at the focal plane along the vertical axis. This property can be utilized to effectively subtract all instrumental background that results from \nuc{6}{Li} scattering elsewhere along the beam line. More specifically, ``true'' \nuc{6}{Li}' events indicative of a scattering off the target will be peaked along the center of the vertical plane, whereas \nuc{6}{Li}$'$ that have scattered elsewhere along the beam line will lie in the tails of this peak. The off-median regions of the vertical focal plane allow for an estimation of the proportion of events within the main peak that are contributions from the instrumental background, which allows for an estimation and subtraction of these events from the spectrum prior to further analysis. 

\begin{figure}[t]
\centering
\includegraphics[width=8.6cm]{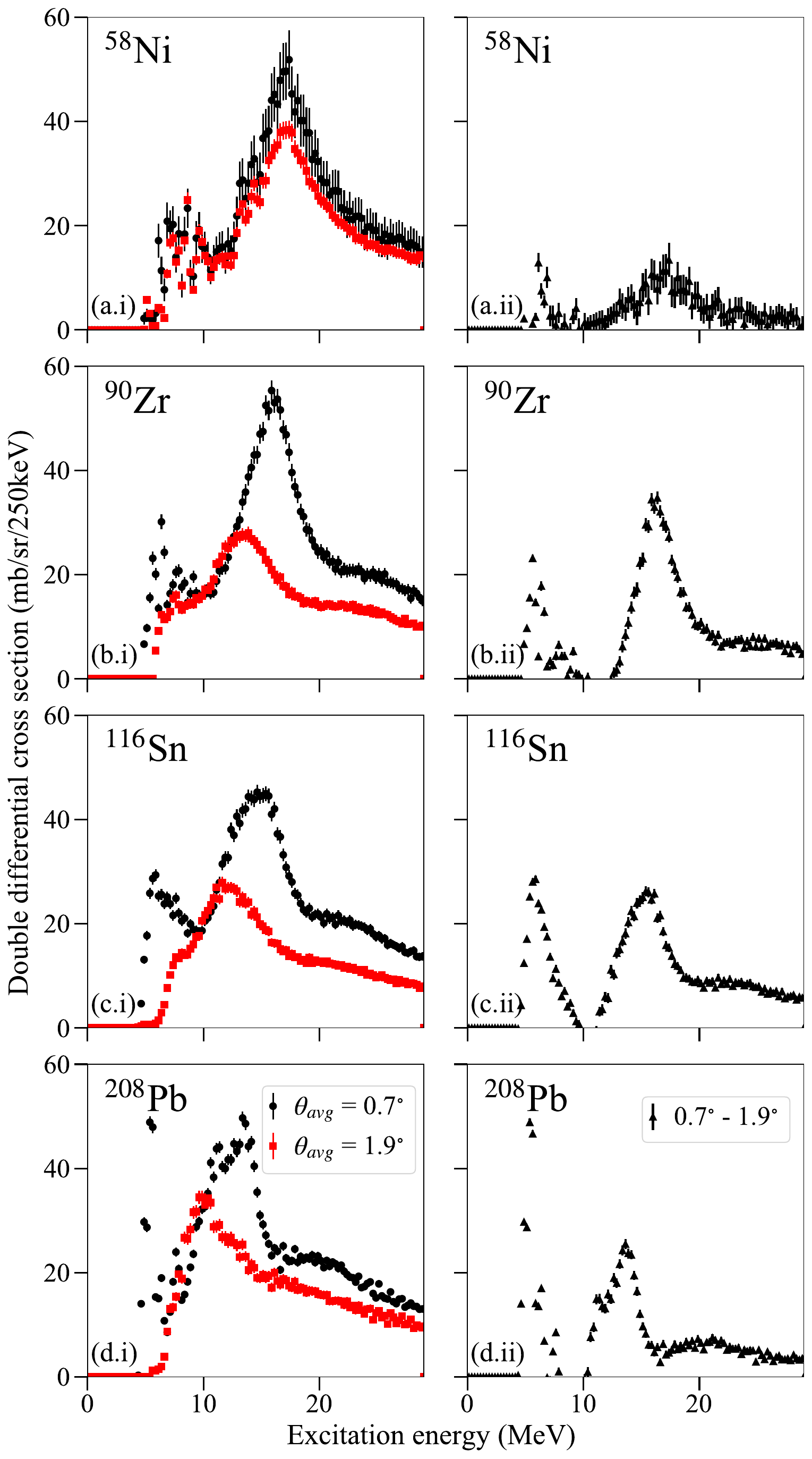}
\caption{\label{fig:fwd_spectra}Background subtracted, extracted inelastic cross sections for all presented nuclei. Shown are the more forward spectra at 0$^\circ$ (0.7$^\circ$ average spectrographic angle) (data points in black) and 1.9$^\circ$ (data points in red) in sub figures (i). Figures on the right, sub figures (ii), are the respective subtraction spectra, showing a strong ISGMR response in each case. The part of the spectrum at the lowest excitation energies likely reflects a potential misplacement of the elastic blocker, resulting in some low-lying states appearing prominently in some inelastic spectra.}
\end{figure}

Elastic scattering as well excitation of low-lying states of \nuc{12}{C} were employed for energy calibration purposes. These discrete states in the horizontal focal plane were fit to Gaussians whose locations yield a correspondence between focal-plane position and ejectile exit channel energy, and thus scattering momentum. With focal plane position and ejectile momentum, relativistic kinematics calculations were performed to generate excitation energy spectra for each target. These 
spectra were binned by angle and energy, and cross sections were calculated from counts in each respective bin. Figure \ref{fig:fwd_spectra} shows forward angle spectra following this data reduction and cross section calculation. Also shown in Fig. \ref{fig:fwd_spectra} (right panels) are subtraction spectra, whereby inelastic scattering spectra at 1.9$^\circ$ are subtracted from those at 0.7$^\circ$. As the angular distribution for the ISGMR is at a minimum at the former angle, as indicated by distorted-wave Born approximation (DWBA) calculations, whereas the isoscalar giant quadrupole resonance (ISGQR) angular distributions are nearly constant over the 0$^\circ$--2.5$^\circ$ region (see Fig. \ref{fig:dwba}), this ``subtraction of spectra'' renders a very good representation of the ISGMR strength distribution \cite{BRANDENBURG198729}. This procedure was used in the present work to verify the quality and veracity of the data prior to the MDA.

\begin{figure*}[!h!t]
    \centering
    \includegraphics[width=17.2cm]{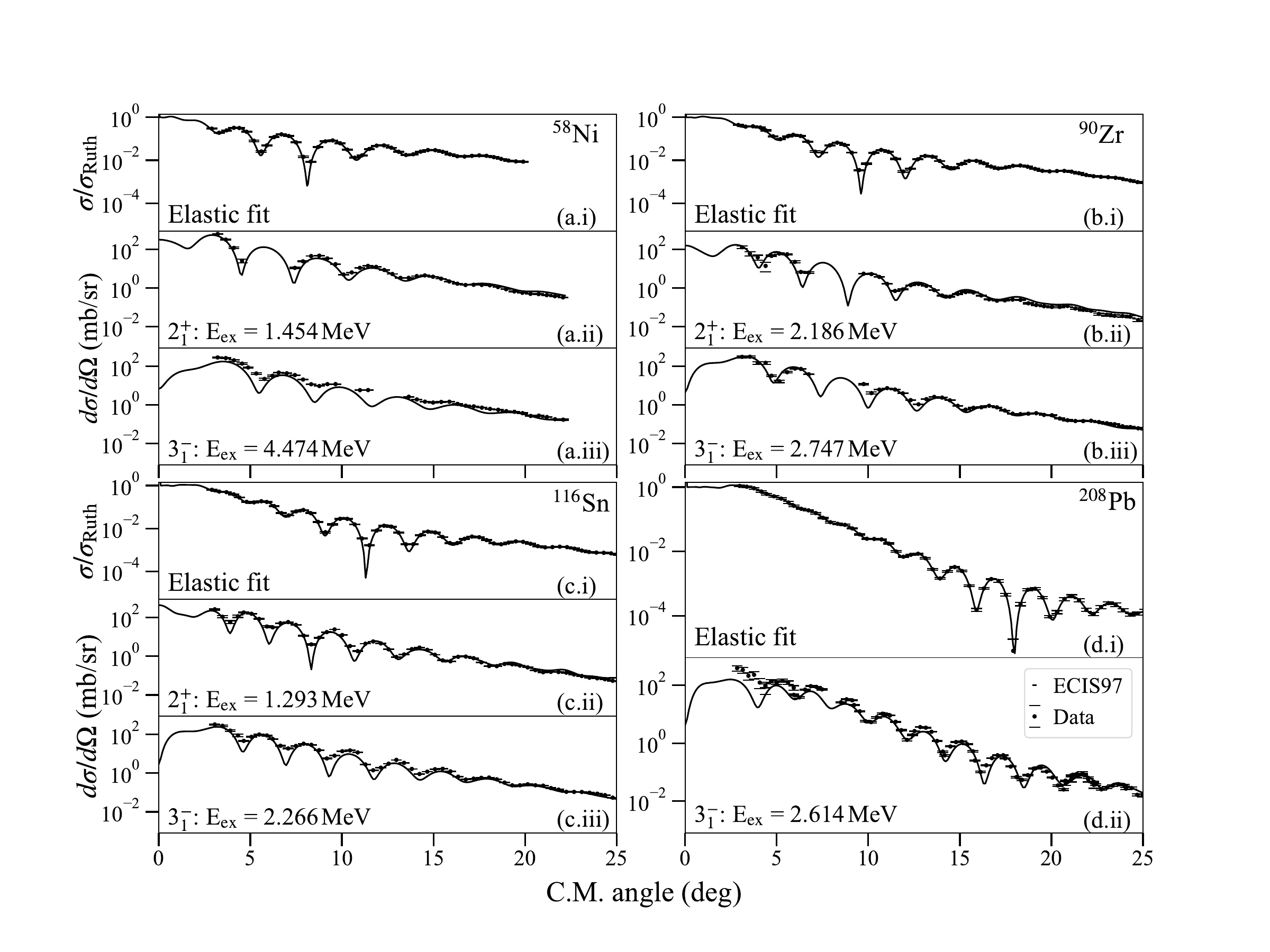}

    \caption{Optical model fits (solid lines) to elastic scattering data for all nuclei are shown in figures (a,i-d,i). Experimental data (squares) for the elastic cross sections are shown in panels labeled with (i). DWBA calculations (solid lines) for low-lying states (shown in squares) using the extracted OMP values are shown in panels labeled with (ii,iii) for \nuc{58}{Ni}, \nuc{90}{Zr}, \nuc{116}{Sn}, and in panel (d,ii) for \nuc{208}{Pb}.}
    \label{fig:OMPs}
\end{figure*}

The optical model parameters (OMP) were extracted using the nuclear scattering code ECIS97 \cite{ecis} and its $\chi^2$ minimization fit routine on the measured elastic angular distributions. Previous efforts to characterize scattering interactions with \nuc{6}{Li} employed hybrid folding and phenomenological optical potentials for real and imaginary terms, respectively \cite{Krishichayan40Ca48Ca,Krishichayan58Ni90Zr,Chen2008}. The optical model in this work utilized a real density-dependent folded potential, imaginary Woods-Saxon volume and surface potentials, and a Coulomb potential. The total optical model is the sum of all these terms, 
\begin{equation}
    U(r) = N_RV_{\mathrm{R,DF}}(r) - i \Big[ W_{\mathrm{Vol}}(r) + W_{\mathrm{Surf}}(r) \Big] + V_{\mathrm{Coul}}(r)
\end{equation}
Here, the Coulomb potential is the standard interaction between a point and uniform sphere. Both the absorptive volume and surface (derivative) potentials utilized the Woods-Saxon form:
\begin{equation}
    \begin{aligned}
        W_{Vol}(r) & = W_Vf(r,R_V,a_V,) \\
        W_{Surf}(r) & = 4a_SW_S \frac{\mathrm{d}}{\mathrm{d}r}f(r,R_S,a_S) \\
        f(r,R_{V,S},a_{V,S}) & = \left[1+\exp{\frac{r-R_{V,S}}{a_{V,S}}}\right]^{-1}.
    \end{aligned}  
\end{equation}

For the real volume potential, the folded approach utilizes the ground-state density distributions of the target and projectile, $\rho_1(\mathbf{r}_1)$ and $\rho_2(\mathbf{r}_2)$ \cite{KhoaDF},

\begin{equation}
    V_{\mathrm{R,DF}}(r) = \int \rho_1(\mathbf{r}_1) \rho_2(\mathbf{r}_2) v(\mathbf{r}_{12}) \mathrm{d}^3r_1 \mathrm{d}^3r_2.
\end{equation}
\noindent The ground state density distribution for \nuc{6}{Li} was extracted from proton scattering experiments \cite{KORSHENINNIKOV199745}. The target densities are two-parameter Fermi distributions obtained from Refs. \cite{PhysRevC.63.034007,SATCHLER1987215}, and were fixed during the OMP search procedure.

For the nucleon-nucleon interaction, the density dependent CDM3Y6-Paris interaction was chosen, following Ref. \cite{Chen2008}. This model assumes that the interaction is energy-dependent, density-dependent, and has the separable form \cite{KhoaDF}:

\begin{equation}
    v_{D(EX)}(\rho, E, \mathbf{s}) = g(E)F(\rho)v^{D(EX)}(\mathbf{s}),
\end{equation}
where $\mathbf{s} = \mathbf{r}_2 - \mathbf{r}_1 + \mathbf{r}$ depends only on the folding coordinates, $g(E)\approx1-0.003\frac{E}{A}$ is an energy dependent factor, $F(\rho) = C[1+\alpha \exp{(-\beta\rho)} - \gamma\rho]$ is a density dependent function, and $v_{D(EX)}$ is the direct (exchange) radial interaction between nucleons. CDM3Y6-Paris interaction parameters can be found in Ref. \cite{KhoatKinf}. The explicit form of $v_{D(EX)}$ is the sum of two (three) phenomenological Yukawa potentials fitted to the oscillator basis reaction-matrix elements \cite{ParisNN}.

The surface term was found to be beneficial in constraining the imaginary volume potential during $\chi^2$ minimization. The omission of a spin-orbit potential follows the work of Refs. \cite{Chen2008,Chen24Mg28Si,SchwandtOMP} and is justified as its effect is minimal \cite{SATCHLER1994241}.

The OMP were arrived at by carrying out systematic parameter sweeps for each set of potentials until $\chi^2$s  for fitting the elastic scattering angular distribution data were no longer affected. The resulting OMP and density parameters are listed in Table \ref{tab:table2}. Shown in Fig. \ref{fig:OMPs} are the fits to the elastic scattering angular distributions for each target nucleus.  The appropriateness of the extracted OMP was tested by calculating the angular distributions for the low lying 2$_1^+$ and 3$_1^-$ states, also shown in Fig. \ref{fig:OMPs}, using the accepted values for the $B(E\mathrm{2})$ and $B(E\mathrm{3})$ transition probabilities \cite{KIBEDI200235,RAMAN20011}.

\setlength{\tabcolsep}{6pt} 
\renewcommand{\arraystretch}{1.1} 
\begin{table*}[]\centering
\caption{Listed are Fermi density parameters from Refs \cite{PhysRevC.63.034007,SATCHLER1987215}, real potential normalization, and Woods-Saxon imaginary volume and surface potentials for all studied targets. Nuclear half-mass radii, $c$, and diffuseness, $a$, were fixed parameters and not varied during the OMP searches. Here, $N_R$ is the normalization of the double-folded, real potential; $W_{V}$, $R_{V}$, and $a_{V}$ are, the depth, radius, and diffuseness of the imaginary volume potential, respectively; and $W_{S}$, $R_{S}$, and $a_{S}$ are the depth, radius, and diffuseness of the imaginary surface potential, respectively. The Fermi density parameters were not part of the fits.
\label{tab:table2}}
\begin{ruledtabular}
\begin{tabular}{cccccccccccc}
 \multirow{2}{*}{Target}&\multicolumn{3}{c}{Density Parameters}&\multicolumn{7}{c}{Optical Model Parameters}\\ \cline{2-11}
 &$\rho_0$& $c$ & $a$ & $N_R$ & $W_V$ & $R_V$ & $a_V$ & $W_{S}$ & $R_{S}$ & $a_{S}$\\
 & &(fm) & (fm) &  & (MeV) & (fm) & (fm) & (MeV) & (fm) & (fm)\\
 \cline{2-11}
 \nuc{58}{Ni} & 0.176 & 4.08 & 0.515 & 0.626 & 68.02 & 4.25 & 0.79 & 4.50 & 6.85 & 0.81\\
 \nuc{90}{Zr} & 0.165 & 4.90 & 0.515 & 0.657 & 87.90 & 4.45 & 0.99 & 5.36 & 7.48 & 0.85\\
 \nuc{116}{Sn} & 0.159 & 5.43 & 0.523 & 0.658 & 106.2 & 4.30 & 1.19 & 5.23 & 7.83 & 0.82\\
 \nuc{208}{Pb} & 0.157 & 6.67 & 0.545 & 0.679 & 44.20 & 6.66 & 0.60 & 13.7 & 8.75 & 0.93\\
\end{tabular}
\end{ruledtabular}
\end{table*}

DWBA calculations were performed using the ECIS97
code to calculate the predicted angular distribution corresponding to 100\% of the energy-weighted sum rule (EWSR) for the various resonances expected to be excited in these measurements. This was done sequentially in 1-MeV excitation energy steps over the entire experimental energy range for all nuclei. Density deformation parameters were calculated externally using the form factors from \cite{Harakeh}. Transition potentials were calculated externally from the optical model utilizing the Tassie model for compression mode oscillations where $\lambda \leq 1$. Further details, as well as relevant compression-mode form factors, can be found in Refs. \cite{Harakeh,UGarg}. Figure \ref{fig:dwba} shows the results from representative DWBA calculations for energies relevant for maximal ISGMR strength response. 

\begin{figure*}[]
    \centering
    \includegraphics[width=17.2cm]{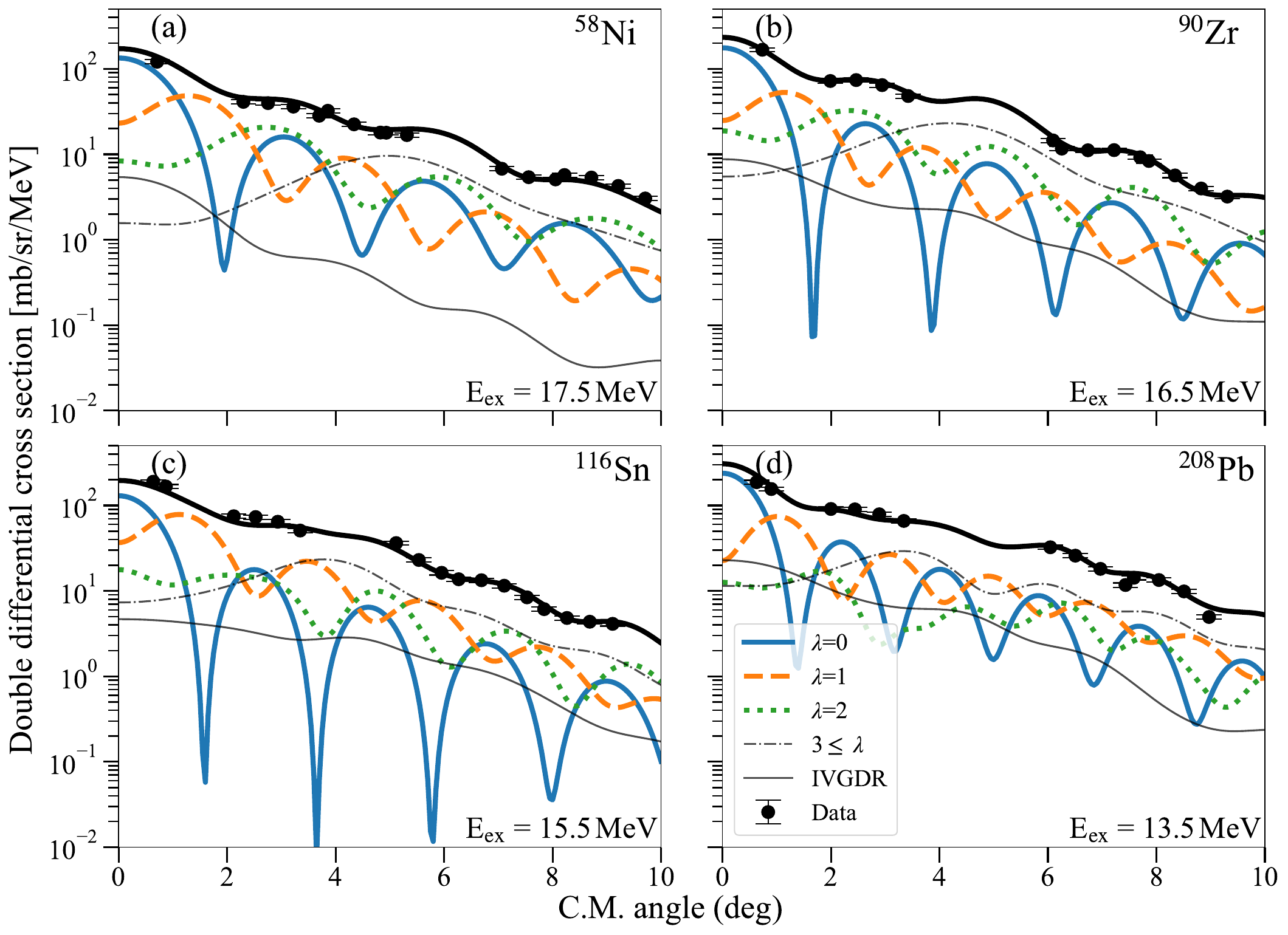}
    \caption{Sample MDA results for excitation energies near ISGMR centroid locations. Shown are the lowest multipolarities, the continuum fit from higher multipolarities, and strength-scaled IVGDR responses. The visible gaps in the angular distribution data are because of unavoidable hydrogen contamination in the target chamber}
    \label{fig:dwba}
\end{figure*}

The calculated DWBA cross sections were used as input in the MDA. This utilized an affine invariant Markov-Chain Monte Carlo (MCMC) algorithm via emcee \cite{2013PASP..125..306F}. The experimental inelastic scattering angular distribution for each energy bin is fitted to a sum of several DWBA cross sections calculated for various multipoles,
\begin{small}
\begin{equation}
    \hspace{-0.2cm} \frac{\mathrm{d}^2\sigma^{exp}(\theta_{C.M.},E_{x})}{\mathrm{d}\Omega \mathrm{d}E} = \sum_{\lambda = 0}^{\lambda_{max}}a_{\lambda}(E_{x}) \frac{\mathrm{d}^2\sigma^{DWBA}_{\lambda}(\theta_{C.M.},E_{x})}{\mathrm{d}\Omega \mathrm{d}E}. \label{eq:6}
\end{equation}
\end{small}

\noindent The ansatz that the experimental angular distributions of the inelastically scattered ejectiles may be modeled purely by a combination of angular distributions corresponding to giant resonance multipoles is the basis for the MDA, whereby the MCMC maximizes the likelihood distribution of the $\chi^2$ by exploring the $\lambda_\mathrm{max}$-dimensional parameter space. More explicit details on this fitting approach can be found in Ref. \cite{2013PASP..125..306F}. 

No significant changes to $\chi^2$s occurred beyond $\lambda_\mathrm{max} \geq 7$. Multipolarities beyond $\lambda \geq 3$ are included to account for the nuclear continuum, which may also contain knock-out/exchange reactions that are not indicative of pure multipoles; these coefficients have been assumed to be non-physical. The contribution of the IVGDR in the experimental distributions was estimated using its adopted strength distributions extracted from photonuclear cross sections  \cite{IVGDR}. The ratio of the strength present in a particular energy bin to the total strength across all excitation energies was used to estimate the IVGDR percentage, $a_\mathrm{IVGDR}$. ECIS97-generated cross sections were then scaled by $a_\mathrm{IVGDR}$ and subtracted out of the experimental distributions prior to model-space exploration in MCMC. Notable MDA results at energies near the peak of ISGMR for each target are shown in Fig. \ref{fig:dwba}.

The MDA yields the percentage of the EWSR exhausted for each multipolarity as a function of excitation energy (\%EWSR). With this, calculation of the ISGMR strengths at each excitation energy from these \%EWSR fractions, $a_0(E_{x})$, results in the ISGMR strength distribution \cite{UGarg,Harakeh}:  

\begin{equation}
    S_0(E_{x}) = \frac{2 \hbar^2 A \langle r^2 \rangle}{m E_{x}} a_0(E_{x}), \label{eq:7}
\end{equation}

\noindent where $m$ is the nucleon mass, $A$ is the mass number of the target, and  $\langle r^2 \rangle$ is the mean-square radius of the ground-state density distribution calculated using a Woods-Saxon form. With the ISGMR strength distribution, moment ratios can be calculated, where the $k$-th moment ratio is defined as 
\begin{equation}
    m_k = \int \mathrm{d}E_{x}S_0(E_{x})E_{x}^k.
\end{equation}
These can be used to characterize the energy of the ISGMR as,
\begin{equation}
    \begin{aligned}
        E_\text{constrained} & = \sqrt{\frac{m_1}{m_{-1}}} \\
        E_\text{centroid} & = \frac{m_1}{m_0} \\
        E_\text{scaling} & = \sqrt{\frac{m_3}{m_1}}.
    \end{aligned}
\end{equation}
In addition to this, a Lorentzian can be fitted to the ISGMR strength distribution if the resonance profile is non-fragmented. This being the case for these medium-heavy nuclei, the ISGMR strength distributions were fitted to Lorentzians of the form:
\begin{equation}
L(E_{x}, I_0, E_{\mathrm{center}}, \Gamma) =I_0\frac{\Gamma}{(E_{x}-E_{\mathrm{center}})^2 + \Gamma^2}. 
\end{equation}

\section*{IV. Results and Discussion}
\begin{figure*}[]
    \centering
    \includegraphics[width=16cm]{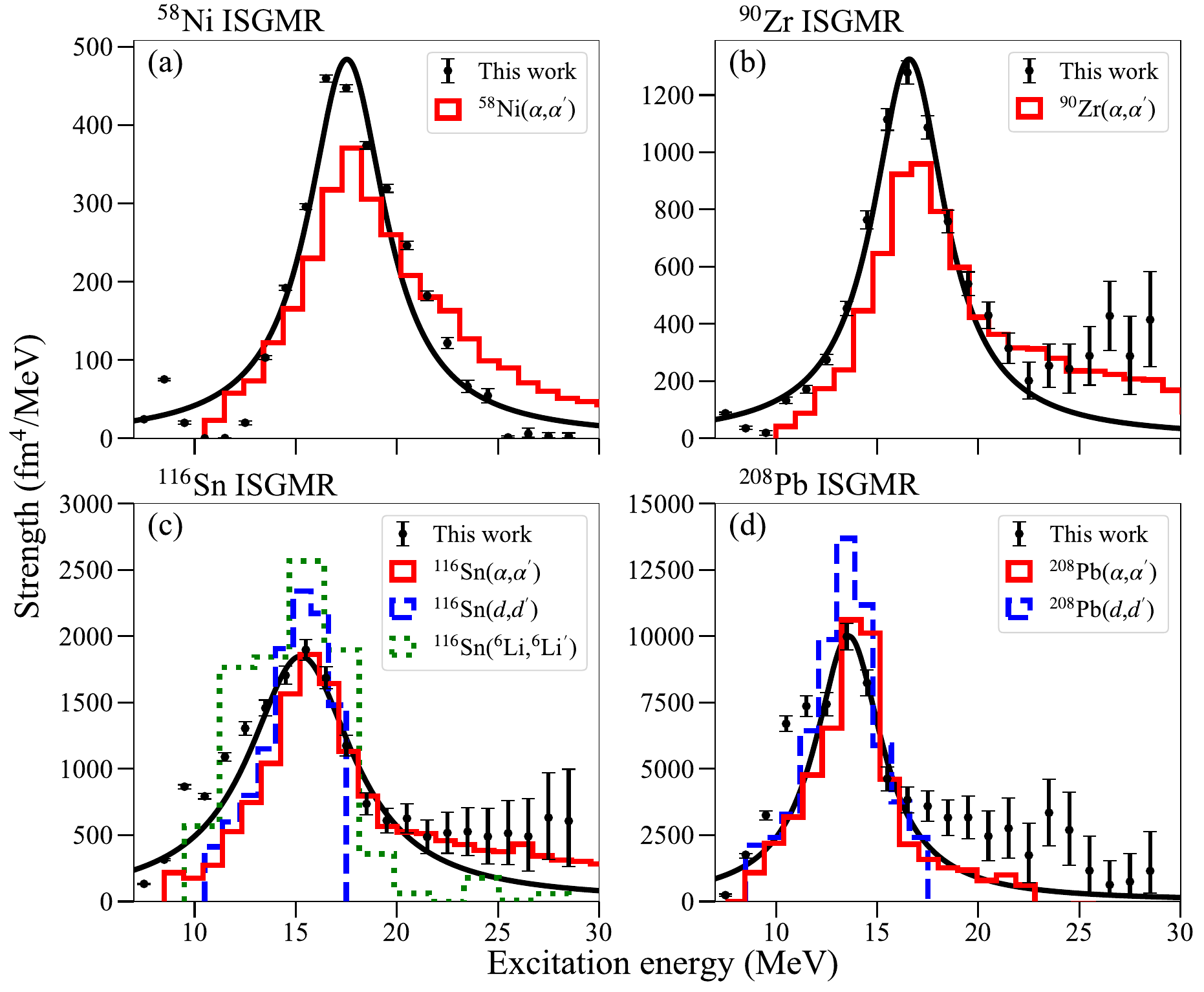}
    \caption{Extracted ISGMR strength distributions for nuclei investigated in this work along with the associated Lorentzian fits (solid lines). Also shown as histograms are strength distributions utilizing $\alpha$ (red, solid), $d$ (blue, dashed), and \nuc{6}{Li} (green, dotted) probes where available for  \nuc{58}{Ni} (a) \cite{BKNayak} ,  \nuc{90}{Zr} (b) \cite{Yogesh90Zr}, \nuc{116}{Sn} (c) \cite{TaoLi_116Sn, DarshanaDeuteron, Chen2008}, and \nuc{208}{Pb} (d) \cite{DarshanaDeuteron, PATEL2013178}.} 
    
    \label{fig:ISGMR_strs}

\end{figure*}

Table \ref{tab:table3} shows centroid energies resulting from Lorenztian fits to the ISGMR strength distributions, as well as calculated moment ratios. Lorentzian fits were performed over the primary resonance response in each strength distribution, typically between 10.5-24.5\,MeV. Total exhausted \%EWSR were calculated from the raw distributions themselves, near the bulk of the resonances, as elaborated upon in Table \ref{tab:table3}. Figure \ref{fig:ISGMR_strs} shows the resulting ISGMR strength distributions from the MDA, and the corresponding Lorenztian fits. Also shown, for comparison, are the ISGMR strength distributions from previous measurements, where available, using $\alpha$-particles and other probes. The error bars here represent 95\% confidence intervals purely from fitting the data and with statistical uncertainties, and do not include other systematic effects resulting from the DWBA, which can be as much as $\pm$20\% for the \%EWSR values \cite{Yogesh90Zr}. The ``extra'' strength at higher excitation energies is spurious and is a consequence of possible limitations of the MDA procedure, as discussed in some detail in Ref. \cite{UGarg}. The presence of this spurious strength also leads to larger values for the extracted strengths (in terms of \%EWSR), depending on the upper end of the range chosen for the purpose. 

\subsection*{\nuc{58}{Ni}}

There is good qualitative agreement with the strength extracted for this nucleus by Nayak et al. \cite{BKNayak}, as seen in Fig. \ref{fig:ISGMR_strs}a. The listed \%EWSRs are generally consistent with our results. Lui et al. \cite{YWLui_58Ni} reported two \%EWSRs for different energy ranges: 58$\pm$6\% between 12-25\,MeV and 74$^{+22}_{-12}$\% between 12-31\,MeV. More relevant here is the former, which better matches the energy range in this work and has similar \%EWSR. Both  previous results \cite{BKNayak,YWLui_58Ni} report similarly higher first moment ratios (19.9$^{+0.7}_{-0.8}$, 20.30$^{+1.69}_{-0.14}$\,MeV respectively). This discrepancy could be attributed to the energy range being limited in this work to a maximum of 28.5\,MeV, resulting in a lower value for this moment ratio.

\subsection*{\nuc{90}{Zr}}

The ISGMR strength distribution and centroid for this nucleus in the present work are in excellent agreement with results from the inelastic $\alpha$ scattering work of Gupta et al. \cite{Yogesh90Zr}. The two strength distributions in Fig. \ref{fig:ISGMR_strs}b appear to be nearly identical, except for the extraneous additional strength beyond 25\,MeV in this work. Nearly all of the ISGMR strength appears to be exhausted in this work, whereas Gupta et al. obtained 74.7$\pm$9\% over 10.5-24.5\,MeV \cite{Yogesh90Zr}. This could be the result of using the raw distribution over a more limited energy range, rather than the Lorentzians themselves; however, the Lorentzian approach has the benefit of disregarding extraneous continuum strength that may not truly be ISGMR and could thus be a more valid approach. The moment ratios reported by Gupta et al. \cite{Yogesh90Zr} are systematically higher than those reported here, although their data range from 10-34\,MeV, which potentially explains this slight discrepancy. While not a direct comparison with this, it is noteworthy that   the 16.8$\pm0.1$\,MeV ISGMR centroid reported by Zamora et al. in the nearby nucleus \nuc{93}{Nb} using a \nuc{6}{Li} probe \cite{Zamora} is also consistent with these results.

\begin{table*}[]
\caption{Lorentzian fit parameters and various moment ratios for all target nuclei. Fits were performed where the bulk of the strength was present, near the centroids. To disregard extraneous continuum strength, \%EWSR was calculated using the resulting Lorentzian fits over the energy range of 10.5-28.5\,MeV. Moment ratios were calculated with the raw distributions and over energy ranges similar to those in the previous work to provide more direct comparisons. Asterisks indicate that energy centers from fits and corresponding widths were obtained using a Gaussian fit over a region containing the bulk of the strength distribution, rather than a Lorentzian as in this work. \label{tab:table3}}
\begin{ruledtabular}
\begin{tabular}{ccccccccc}
 \multirow{2}{*}{Target} & \multirow{2}{*}{Probe[Ref]} &  & $E_{\mathrm{center}} $ & $\Gamma$ & $\frac{m_1}{m_0}$ & $\sqrt{\frac{m_1}{m_{-1}}}$ & $\sqrt{\frac{m_3}{m_1}}$ & \multirow{2}{*}{\%EWSR} \\
 & & & (MeV) & (MeV) & (MeV) & (MeV) & (MeV) \\  \hline 
 \nuc{58}{Ni} & This work & & 17.5$\pm$0.1 & 4.6$\pm$0.1 & 17.8$\pm0.1$  & 17.4$\pm$0.1 & 18.6$\pm0.1$ & 79$\pm$2 \\
 \nuc{58}{Ni} & $\alpha\text{\cite{BKNayak}}$ & & - & - & 19.9$^{+0.7}_{-0.8}$ & - & - & 92$^{+3}_{-4}$ \\
 \nuc{58}{Ni} & $\alpha\text{\cite{YWLui_58Ni}}$ & & - & 4.25$^{+0.69*}_{-0.23}$ & 20.30$^{+1.69}_{-0.14}$ & 19.93$^{+1.34}_{-0.07}$ & 21.43$^{+3.01}_{-0.32}$ & 58$\pm6$ \\
 \nuc{90}{Zr} & This work & & 16.6$\pm$0.1 & 4.4$\pm$0.1 &18.5$^{+0.5}_{-0.6}$ & 17.9$^{+0.4}_{-0.5}$ & 20.3$^{+0.6}_{-0.8}$ & 95$\pm$2 \\
 \nuc{90}{Zr} & $\alpha\text{\cite{Yogesh90Zr}}$ & & 16.76$\pm$0.12 & 4.96$^{+0.31}_{-0.32}$ & 19.17$^{+0.21}_{-0.20}$ & 18.65$\pm$0.17 & 20.87$^{+0.34}_{-0.33}$ & 74.7$\pm$9 \\
 \nuc{116}{Sn} & This work & & 15.3$\pm$0.2 & 6.3$\pm$0.2 & 15.3$\pm$0.1 & 15.1$\pm$0.1 & 16.1$\pm$0.1& 112$^{+9}_{-8}$\\
 \nuc{116}{Sn} & $\alpha\text{\cite{TaoLi_116Sn}}$ & & 15.8$\pm$0.1 & 4.1$\pm$0.3 & 15.8$\pm$0.1 & 15.7$\pm$0.1 & 16.3$\pm$0.2 & 86$\pm$5 \\
 \nuc{116}{Sn} & $d\text{\cite{DarshanaDeuteron}}$ & & 15.7$\pm$0.1 & 4.6$\pm$0.7 & - & - & - & 73$\pm$15 \\
 \nuc{116}{Sn} & \nuc{6}{Li}$\text{\cite{Chen2008}}$ & & 15.58$\pm$0.19$^{\text{*}}$ & 5.46$\pm$.18${\text{*}}$ & 15.39$^{+0.35}_{-0.20}$ & - & - & 106$^{+27}_{-11}$ \\
 \nuc{208}{Pb} & This work & & 13.6$\pm$0.1 & 4.0 $\pm$0.2 & 12.8$\pm$0.1 & 12.6$\pm$0.1 & 13.3$\pm$0.1 & 143$^{+9}_{-11}$ \\
 \nuc{208}{Pb}  & $\alpha\text{\cite{PATEL2013178}}$ & & 13.7$\pm$0.1 & 3.3$\pm$0.2 & 13.7$\pm$0.1 & - & - & \\
 \nuc{208}{Pb}  & $\alpha\text{\cite{UCHIDA200312}}$ & & 13.5$\pm$0.2* & 3.6$\pm$0.4 & 13.5$\pm$0.2 & - & - & 76$\pm$5 \\
 \nuc{208}{Pb}  & $d\text{\cite{DarshanaDeuteron}}$ & & 13.6$\pm$0.1 & 3.1$\pm$0.4 & - & - & - & 147$\pm$18 \\
\end{tabular}
\end{ruledtabular}
\end{table*}

\subsection*{\nuc{116}{Sn}}

There is a plethora of data for comparison in this nucleus. The ISGMR strength distribution  reported by Li et al. \cite{TaoLi_116Sn} from inelastic $\alpha$ scattering overlaps quite well with the ISGMR strength distribution presented here. Those efforts resulted in an $E_{\mathrm{center}}$ of 15.8$\pm$0.1\,MeV. The slight discrepancy may be because of a stronger ISGMR strength response at lower energies in the present work. \%EWSRs within the main bulk of the strength distributions are still consistent with each other over a similar energy range. 

Results from $d$ \cite{DarshanaDeuteron} and \nuc{6}{Li} \cite{Chen2008} inelastic scattering  are also shown for comparison. The $d$ data is somewhat limited, but the strength distribution is overall quite similar. As well, the strength distribution reported in the previous \nuc{6}{Li} measurement \cite{Chen2008} matches quite well with the present results. The data points shown in Fig. 4 from Ref. \cite{Chen2008} were converted from \%EWSR data presented therein, and scaled with relevant form factors. The peak-fit parameters reported in Ref. \cite{Chen2008} utilized a Gaussian parameterization, and thus are not as useful for a direct comparison. There is a good similarity between the strength distribution from Ref. \cite{Chen2008} and the one presented here, despite a different experimental setup and subtraction of an empirical background in their analysis. The \%EWSR of this work is in agreement with previous $d$ and \nuc{6}{Li} data.

\subsection*{\nuc{208}{Pb}}

The results from the present work are quite consistent with those from inelastic $\alpha$ scattering \cite{PATEL2013178}, 
with an $E_{\mathrm{center}}$ that is generally in agreement with the others. The \%EWSR is much larger than previously reported by Uchida et al. \cite{UCHIDA200312}, despite their ISGMR strength distribution appearing quite similar to that in Ref. \cite{PATEL2013178} and that reported here. It is quite possible that this is due to the somewhat poor quality of the data for this nucleus in the present work. Some of it might also be a consequence of utilization of the Lorentzian for the calculation of \%EWSR in this work, rather than the raw distributions.

Also available for comparison are results from $d$-scattering \cite{DarshanaDeuteron}. These data have a limited energy range, but still match quite well with the strength distribution obtained in the present work, with a good agreement also for 
$E_\mathrm{center}$ of 13.6$\pm$0.1\,MeV. The \%EWSR value reported in Ref. \cite{DarshanaDeuteron}, 147$\pm$18 \%, is much closer to the present results.

There appears to be some extraneous ISGMR strength at 10.5-11.5\,MeV. This is possibly due to the MDA failing to fully distinguish the ISGQR from the ISGMR. This could be the result of the lack of intermediate angle data that could not be utilized due to the presence of Hydrogen contaminant. Another explanation could be the lack of 1.2-1.7$^{\circ}$ data near the first minimum of the ISGMR angular distribution (see Fig. \ref{fig:dwba}), as its presence would provide more effective discriminatory data points in the MDA $\chi^2$ minimization.

\section*{V. Conclusion}

This work has attempted to establish the viability of using small-angle \nuc{6}{Li} scattering in the investigations of the ISGMR via direct comparison with previously performed experiments utilizing the now ``standard'' $\alpha$ probes. Overall, ISGMR centroid extractions for all nuclei are consistent with previous measurements. The only slight exceptions to this statement are somewhat lower $E_\mathrm{center}$ for \nuc{58}{Ni} and \nuc{116}{Sn} in this work; however, the strength distributions themselves match very well with those available with the probes utilized previously. The agreement with previous results may be further improved if the \nuc{6}{Li} measurements also made at an energy similar to that used in the $\alpha$ and $d$ measurements used for comparison {\em viz.} 100\,MeV/u. As these investigations move towards inverse kinematics reactions with radioactive ion beams to probe ISGMRs in nuclei far from the stability line, utilization of \nuc{6}{Li} scattering might be a valuable complement to the work with active targets.

\section*{Acknowledgments}

We acknowledge the RCNP cyclotron staff for their efforts in providing the high-quality \nuc{6}{Li} beams required for these measurements and thank H. Hashimoto (RCNP) and K. Sault (Notre Dame) for their help with the experiment. This work has been supported in part by the U. S. National Science Foundation (Grants No. PHY-0822648, No. PHY-1068192, No. PHY-2011890 and No. PHY-2310059).

\medskip

\pagebreak

\end{document}